\documentclass[conference]{IEEEtran}
\pdfoutput=1
\usepackage{cite}
\usepackage{amsmath,amssymb,amsfonts}
\usepackage{graphicx}
\usepackage{textcomp}
\usepackage{xcolor}
\usepackage{float}
\usepackage{subcaption}
\usepackage{algorithm}
\usepackage{algpseudocode}
\usepackage{pgfplots}
\usepackage{float}
\pgfplotsset{
	group size/.initial=3,
	group gap/.initial=0.01cm,
}
\makeatletter
\renewcommand{\ALG@beginalgorithmic}{\small}
\makeatother
\usepackage{caption}
\def\BibTeX{{\rm B\kern-.05em{\sc i\kern-.025em b}\kern-.08em
    T\kern-.1667em\lower.7ex\hbox{E}\kern-.125emX}}
\begin{document}

\title{Dual Connectivity Support in 5G Networks: An SDN based approach
}

\author{
	\IEEEauthorblockN{Pradnya Kiri Taksande\IEEEauthorrefmark{1},
	Pranav Jha\IEEEauthorrefmark{1},
	Abhay Karandikar\IEEEauthorrefmark{1}\IEEEauthorrefmark{2}}
	\IEEEauthorblockA{\IEEEauthorrefmark{1}Department of Electrical Engineering,
		Indian Institute Technology Bombay, India 400076\\
		Email: \{pragnyakiri,pranavjha,karandi\}@ee.iitb.ac.in}
	\IEEEauthorblockA{\IEEEauthorrefmark{2}Director and Professor, Indian Institute Technology Kanpur, India 208016\\
	Email: karandi@iitk.ac.in
		}
}
\maketitle
\begin{abstract}
Dual Connectivity (DC) is one of the key techniques to harness the potential of heterogeneous cellular networks. However, 3rd Generation Partnership Project (3GPP) has introduced disparate mechanisms for DC support in different Radio Access Technologies (RATs), bringing complexity to the network nodes in a Multi-RAT Radio Access Network (RAN). Moreover, DC support requires an exchange of high volume of control information between these network nodes. To address these issues, in this paper an SDN based architecture for Multi-RAT RAN is proposed. The proposed architecture brings simplicity to the network interactions, allows flexibility in the network and helps the network in performing load balancing and mobility management functions effectively. We also demonstrate a reduction in the control signaling and an improvement in system performance compared to legacy architecture.
\end{abstract}

\section{Introduction}
The massive growth in the usage of mobile devices and gadgets has led to an exponential increase in data traffic in mobile networks. According to an estimate, the global mobile data traffic is expected to increase sevenfold from 2016 to 2021\cite{cisco}. One of the provisions to meet the increased data traffic demands is to bring the network infrastructure closer to the end users, that is, reducing the cell coverage and increasing the density of cells in a given geographical area. The fusion of wide coverage macro cells overlaid with varying coverage Small Cells (SCs) belonging to different Radio Access Technologies (RATs), e.g., Long Term Evolution (LTE), 5G, Wireless Local Area Network (WLAN) (popularly known as Wireless Fidelity (Wi-Fi)) is known as a Heterogeneous Network (HN). The macro cells provide coverage while the SCs yield capacity. This heterogeneous nature of the deployed network, however, leads to an increase in the number of handovers of User Equipments (UEs) between cells. To address these mobility challenges of HN, 3rd Generation Partnership Project (3GPP) has introduced Dual Connectivity (DC) \cite{36842} as a part of LTE Release 12. \par
In DC, a UE is connected to two base stations simultaneously. One of the base stations is called a Master Node (MN) and the other a Secondary Node (SN). The control plane functionality of the UE is handled by the MN, which is typically a macro base station while the user plane functionality can be handled by MN, SN or both. Varied modes of DC have been defined in 3GPP standards \cite{36842},\cite{36300}, \cite{37340}. DC between LTE MN and LTE SN is defined in \cite{36842}. DC between LTE MN and WLAN SN is characterized in \cite{36300} as LTE WLAN Aggregation (LWA). DC between LTE/5G MN and LTE/5G SN is known as Multi-RAT Dual Connectivity (MR-DC) \cite{37340}. The control signaling procedures in these different modes are not uniform and differ in many aspects. For instance, in LTE DC, control information for a UE can be transferred via MN only while in MR-DC, it can be transferred via MN or MN and SN both. Also, MN and SN exchange the control information for a dual connected UE through X2 interface between them. Because of a single control plane connection for a UE to the Core Network (CN) via the MN, the volume of control information to be exchanged between MN and SN is significant. This increases the control load on each of the MNs for DC UEs. Again, this control load increases with the number of DC UEs. Due to these issues, the existing Radio Access Network (RAN) architecture is not favorable for the management of DC UEs in a multi-RAT network. \par
Software-Defined Networking (SDN) is an emerging paradigm for network control and management in which the control plane is separated from the data plane. To simplify the existing architecture and to reduce the control signaling between the network nodes in the RAN, we propose an SDN based architecture for multi-RAT Radio Access Network. With the control and data planes separated, the architecture is greatly simplified, and control signaling between the network nodes reduces considerably. Also, this leads to better load and mobility management as well as flexibility in controlling the network. \par
There are existing works in the literature which exploit the SDN framework in the field of wireless networks. In \cite{sdwn}, the authors introduce a common mobile network controller to control the UEs, RAN and the Core. \cite{systemLevel} presents a software-defined architecture for LTE system modifying the CN as well as RAN. The authors in \cite{hierarchical} propose an SDN architecture consisting of a centralized superior SDN controller and a localized subordinate SDN controller. However, the decisions are distributed between the two controllers and not completely centralized. \par
In \cite{SoftRan}, the authors propose a software-defined RAN consisting of a centralized RAN controller and radio elements. The centralized controller takes decisions that affect the global state of the network. Each radio element takes local control decisions which do not affect the neighboring elements. As opposed to this, we propose a centralized control plane architecture for multi-RAT RAN. In \cite{5gmultirat}, the authors introduce a tight aggregation of Multi-RAT network using a common RAN cloud. However, there is no clear separation of the control plane from the data plane in this work. In contrast to this, our SDN based architecture has a clean separation of the control plane from the data plane which makes it a simplified architecture. In \cite{akshatha}, the authors propose a centralized SDN architecture for 5G by relocating the control functions from RAN to the CN. In \cite{karandikar}, the authors propose a centralized SDN based wireless network controller for controlling the multi-RAT nodes as well as gateways in the core network. However, in our work, we focus on SDN based control of interworking among multiple RATs at the RAN level from the perspective of dual connectivity. \par
The main features of the proposed architecture are as follows: (1) An SDN based RAN architecture with separate control and data plane entities, (2) Support for multi-RAT RAN integration with the help of a logically unified controller, (3) Support for DC including inter-RAT scenario, (4) Reduction in control signaling, (5) Flexible architecture, (6) Improved load and mobility management, and (7) Improved system performance. To the best of our knowledge, this is the first work to present the impact of an SDN based multi-RAT RAN architecture on dual connectivity. \par
The rest of the paper has been organized as follows. In Section \ref{sec2}, we outline different DC architectures as proposed by 3GPP. We describe the proposed architecture in Section \ref{sec3}. In Section \ref{sec4}, we show that there is a significant reduction in the control signaling exchange and improvement in system throughput compared to legacy architecture. We conclude the paper in Section \ref{conc}.\par
\section{Background}
\label{sec2}
The different modes of DC, as defined in 3GPP standards, have been elaborated in this section.
\subsection{LTE DC Architecture}
\label{DC}
\begin{figure}
	\centering
	\begin{subfigure}[]{0.235\textwidth}
		\centering
		\includegraphics[width=\textwidth]{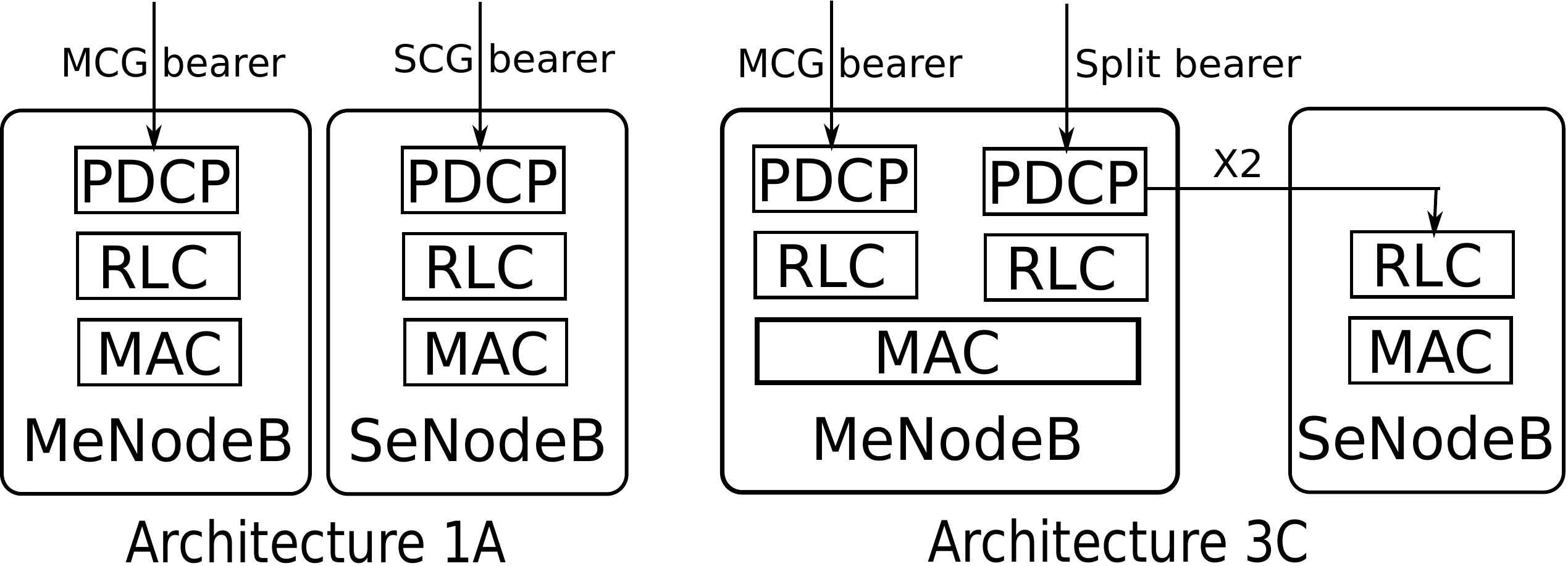}
		\caption{LTE DC Architecture.}
		\label{DCprot}
	\end{subfigure}
	~
	\begin{subfigure}[]{0.235\textwidth}
		\centering
		\includegraphics[height=3cm]{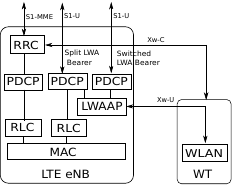}
		\caption{LWA Architecture.}
		\label{LWA-f}
	\end{subfigure} %
	\caption{LTE DC and LWA Architectures.}
\end{figure}
The LTE network is comprised of two components: CN and Evolved Universal Terrestrial Radio Access Network (E-UTRAN). The CN consists of Packet Data Network Gateway (PGW), Serving Gateway (SGW) and Mobility Management Entity (MME) in addition to a few other nodes. The E-UTRAN (or simply RAN) consists of eNodeBs (eNBs) and the radio interface to transmit/receive signals and data to/from UEs. DC architecture between LTE MN and LTE SN is displayed in Figure \ref{DCprot}. In LTE DC, the control plane for a dual connected UE resides in the MN, and its user plane can be handled by MN, SN or both. The MN may be a single cell or comprise of a group of cells called the Master Cell Group (MCG). Similarly, the SN may be a single cell or comprise of a group of cells called the Secondary Cell Group (SCG). The Radio Resource Control (RRC) layer of the eNB (MN as well as SN) performs the radio control functions such as radio bearer set-up, UE measurement reporting, mobility management, the configuration of data plane protocol layers. These details may need to be exchanged between MN and SN via the X2 interface. \par
Radio bearers are pipes created by eNBs for the transfer of either user or control data to/from UEs. The bearers used for the exchange of control information are known as Signaling Radio Bearers (SRBs), and those used for the transfer of user data are known as Data Radio Bearers (DRBs). The data transfer from the PGW to a DC UE takes place via an MCG or SCG or split bearer as illustrated in Figure \ref{DCprot}. In the case of 1A architecture, as defined by 3GPP \cite{36842}, separate bearers are established from PGW to UE via MN (MCG bearer) and SN (SCG bearer). In the case of 3C architecture \cite{36842}, also known as split bearer architecture, a single data bearer from the PGW is split into two radio bearers at the MN, one carrying data via MN and the other carrying data via SN. This split takes place at the Packet Data Convergence Protocol (PDCP) layer of MN, and the data is sent by MN over the X2 interface to SN to be forwarded to the UE. The SCG bearer and split bearer are to be used for data transfer only while the MCG bearer can carry data as well as signaling. \par
\subsection{LTE-WLAN Aggregation (LWA)}
\label{LWA}
Similar to LTE DC, in LTE Release 13 and 14, radio level traffic aggregation over LTE and WLAN has been introduced as LTE-WLAN Aggregation (LWA) \cite{36300} by 3GPP. To allow a WLAN Access Point to take the role of SN, 3GPP has specified a logical node called WLAN Termination (WT). For UEs configured with LWA, a split/switched LWA data bearer, as illustrated in Figure \ref{LWA-f} may be routed via both eNB and WT or WT alone, while the control plane connection remains on LTE eNB. The user data is sent through WT using the LWA Adaptation protocol (LWAAP). A new interface called "Xw" has been defined for the exchange of control and data between eNB and WT.\par
\subsection{Multi-RAT Dual Connectivity}
\label{MR-DC}
Multi-RAT Dual Connectivity (MR-DC) \cite{37340} is a generalization of DC described in Section \ref{DC}, where dual connectivity is established with LTE and 5G radio network for a UE. Either LTE eNB or 5G gNB acts as the MN and SN. MN and SN are connected via an Xn interface, and at least the MN is connected to the CN. In MR-DC, there is a new Signaling Radio Bearer SRB3, which is established between SN and UE. Besides, there is a Split SRB, which is an SRB between the MN and the UE allowing the duplication of control information via MN and SN both. However, MN always sends the initial SN RRC configuration via MCG SRB (SRB1), but succeeding reconfigurations may be carried via MN or SN.\par
\section{Proposed SDN-based 5G multi-RAT architecture}
\label{sec3}
\subsection{Motivation}
\label{motivation}
We first discuss the limitations of the existing 3GPP architectures. In LTE DC, as explained in Section \ref{DC}, there is only one RRC connection, and the SRB (SRB1 and SRB2) is served by the MN only. When SN has to share radio control information with UE, it needs to send this information to the MN over the X2 interface, which in turn sends it to UE. Also, the SN has to get the data bearer related control information from the MN through the X2 interface. This leads to an exchange of a large number of signaling messages between MN and SN for a DC UE. The amount of control information exchange increases with an increase in the number of DC UEs in the network. Besides, this exchange takes place over the non-ideal backhaul link between MN and SN which has a delay of the order of milliseconds leading to additional delay in the signaling exchange. \par
As described in Section \ref{LWA}, in LWA, there is a single UE specific signaling connection between CN and eNB. Also, all control information exchange with UE takes place via eNB, i.e., both SRB1 and SRB2 are carried by eNB. The control information is exchanged between WT and eNB via the Xw-C interface. In MR-DC, there are three types of Signaling radio bearers: SRB1 and SRB2 can be split across both MN and SN, and SRB3 is through SN. The initial signaling, however, can happen through MN only. Thus, the SRB establishment procedure is not uniform across different DC architectures. A significant amount of control information exchange takes place between MN and SN via the non-ideal backhaul link causing additional delay.\par
In the existing architecture, the control function is distributed across the network nodes in RAN. It is difficult to get load information of all the cells at a central location. Hence, the traditional method of connecting a UE to a node having the best signal strength fails to take into account the possibly differing load levels across cells in a heterogeneous environment. Also, the load and interference management information is shared via the X2/Xw/Xn interface between two network nodes for load balancing and interference coordination between them. This sharing, however, typically happens between a small number of neighboring nodes only. Hence, the decisions taken for a DC UE association locally by a network node may not be optimal. \par
Therefore, to simplify the complexity of the existing architecture and to achieve better load balancing in the network, we propose an SDN based multi-RAT RAN architecture, where the control and data plane are separated. As a result, signaling and data flow management are simplified. Also, the data plane nodes are controlled by a centralized controller which has a global view of the load and interference conditions in the network. Through this, the problem of selection of an appropriate MN and SC for a DC UE can be addressed efficiently. 
\subsection{Description of Proposed Architecture}
\label{PropArch}
\begin{figure}
	\centering
	\includegraphics[width=8cm, height=8.5cm]{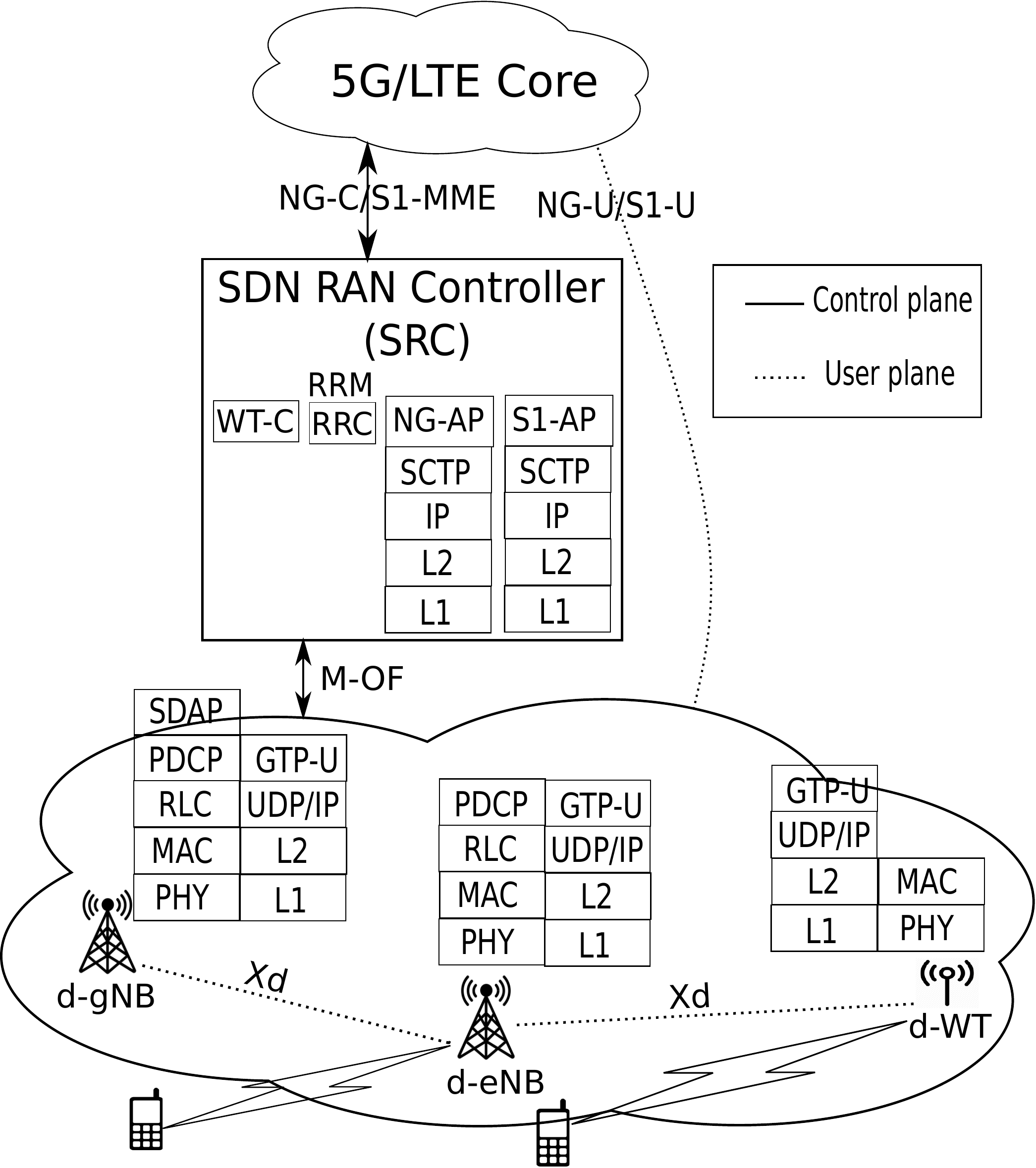}
	\caption{Proposed multi-RAT network architecture.}
	\label{Overproarch}
\end{figure}
In this section, we explain the proposed SDN based RAN architecture for multi-RAT networks as illustrated in Figure \ref{Overproarch}. In this architecture, the Core Network can consist of LTE Core or 5G Core (5GC). The RAN is divided into two parts: i) An SDN RAN Controller (SRC) with the combined control plane functionality of LTE eNBs, 5G gNBs, and WLAN WT points. ii) Diverse data plane entities, also called d-eNBs (data-eNBs), d-gNBs (data-gNBs), and d-WTs (data-WTs); created out of LTE eNBs, 5G gNBs, and WTs, respectively by removing their control functions. The data plane entities such as d-gNBs and d-eNBs can further be divided into Centralized Data-plane Units (CDUs) facing the CN and Distributed Data-plane Units (DDUs) facing UEs. CDUs can act as RAN data plane aggregation entities. \par
In the legacy LTE architecture, each of the eNBs handles the Radio Resource Management (RRM) and control functions for the respective UEs associated with them. However, in the proposed architecture, the RRM functionality, the control plane functionalities and the associated protocol layers of LTE eNB (RRC, RRM, and S1 Application Protocol (S1-AP)) have been transferred to SRC. Similarly, the control functionality of 5G gNB (RRC, RRM, and Next Generation Application Protocol (NG-AP)) and WT (WT-C) has been shifted to SRC. The control functionality of all the nodes is thus, centralized at SRC. There is no control information exchange between the data plane entities. This results in a reduction in the amount of signaling required in DC operations such as cell addition, cell change and cell release as illustrated in Section \ref{sec4.1}. \par
The data plane entities consist of data plane functions as well as the associated protocol layers for communicating with the UE over the air interface. For instance, the protocol layers Service Data Adaptation Protocol (SDAP), PDCP, Radio Link Control (RLC), Medium Access Control (MAC), Physical (PHY) of gNB and PDCP, RLC, MAC, PHY of eNB exist in d-gNB and d-eNB, respectively to exchange data with a UE. Similarly, d-WT hosts the protocol stack to exchange data with a UE over the WLAN interface. The data plane entities exchange data among themselves using GPRS Tunnelling Protocol for User Plane (GTP-U) on top of User Datagram Protocol (UDP). The data exchange between the data plane nodes takes place through the Xd interface, which is a new interface for the transfer of data. Xd is a common interface for communication between data plane nodes of different RATs. Data is exchanged between the CN and data plane entities through the NG-U/S1-U user plane interface using the GTP-U, UDP/IP, L2, L1 protocol stack. \par
SRC uses the Next Generation control plane (NG-C) or S1-MME interface for exchange of control information with the CN depending on whether it is communicating with 5G or LTE CN. SRC controls the data plane entities with a southbound interface using a modified OpenFlow (M-OF) protocol based on OpenFlow (OF) \cite{OF} and OF-Config \cite{OF-C}. SRC has a global view of all data plane entities to take effective decisions. For instance, according to the radio, memory and CPU load information for each of the d-gNB, d-eNB and d-WT at the SRC, an appropriate path for a bearer can be chosen. \par
\subsection{Features of Proposed Architecture}
\label{features}
\subsubsection{Centralized control logic at SRC}
The control functions of all nodes in the RAN reside at the SRC. As a result, the overall management of RAN can be improved. Depending on the QoS requirement of a traffic flow, the data path can be set up through an appropriate data plane node. For instance, for a best effort type traffic, a data path through d-WT can be set up. For a traffic flow with real-time constraints, e.g., for a voice call, a data path through d-eNB could be established. Similarly, for a low latency application, the data path through a d-gNB could be set up. The control functionality at the SRC then conveys their respective configurations to d-WT, d-eNB, and d-gNB individually. Additionally, the architecture helps in administering the entire RAN with the help of simple configuration commands from the controller.
\subsubsection{Flexibility in setting up an SRB}
In legacy LTE architecture, the SRB is usually served by macro eNB. However, in the proposed architecture, the RRC functionality of all data plane nodes exists at the SRC. So there are no specific requirements as to which node should serve the SRB. The SRB can be served by any cell, irrespective of its coverage according to UE's radio conditions. For instance, a SC can serve the SRB for low mobility UEs. Such type of changes may be difficult in the existing architecture.
\subsubsection{Load Management}
SRC has a global view of the load and interference conditions of the different multi-RAT nodes in the network. It is responsible for the control functions of RAN. The data transfer between the network nodes in RAN takes place according to these control functions. These control functions allow the SRC to associate UEs with those nodes which are less loaded especially for the secondary link in case of DC. This leads to higher throughput through better distribution of load across the nodes. We propose a centralized algorithm for DC in Section \ref{sec4.2}, and show that it leads to an improvement in the system performance.
\subsubsection{Mobility Management}
During a handover, the change in the path from a source node to a target node at the CN is known as path switch procedure. In the proposed architecture, in the case of a handover from a source data plane node to a target node, SRC may decide not to perform path switch procedure, in certain scenarios and allow the source data plane node to act as the anchor node for the UE. For example, if the current session of the UE is expected to be a short one, e.g., a Voice over IP (VoIP) session, such a decision can be taken by the SRC. The data proceeds from the source node to the target node and then to the UE in this case. A similar action can be taken in case the source node has data stored in its cache that the UE wants to download. In this situation, updating the path of the UE to the target node may lead to a longer delay in delivering the data to the UE. Such dynamic changes in mobility management procedures by the centralized controller according to session requirement leads to improved network behavior.
\subsubsection{Flexibility in processing the data plane protocol layers across nodes}
The proposed architecture can bring flexibility in the processing of data plane protocol layers across the data plane entities in the RAN. SRC is aware of the load, resources, capabilities and buffer status of the data plane entities. If a UE connected to a macro cell requests for DC with a SC, for instance, the SRC might decide to perform RLC layer processing at the data plane node (d-eNB/d-gNB) of either the macro cell or SC according to the resources and capabilities available at the respective cells.
\subsubsection{Reduction in Control Signaling}
The proposed architecture leads to a reduction in control signaling overhead in the network, thereby improving the performance of the system. Section \ref{sec4.1} demonstrates this reduction in control signaling.
\section{Performance Analysis \& Simulation Results}
\label{sec4}
\subsection{Reduction in control signaling}
\label{sec4.1}
\begin{figure}
	\centering
	\includegraphics[width=8.6cm]{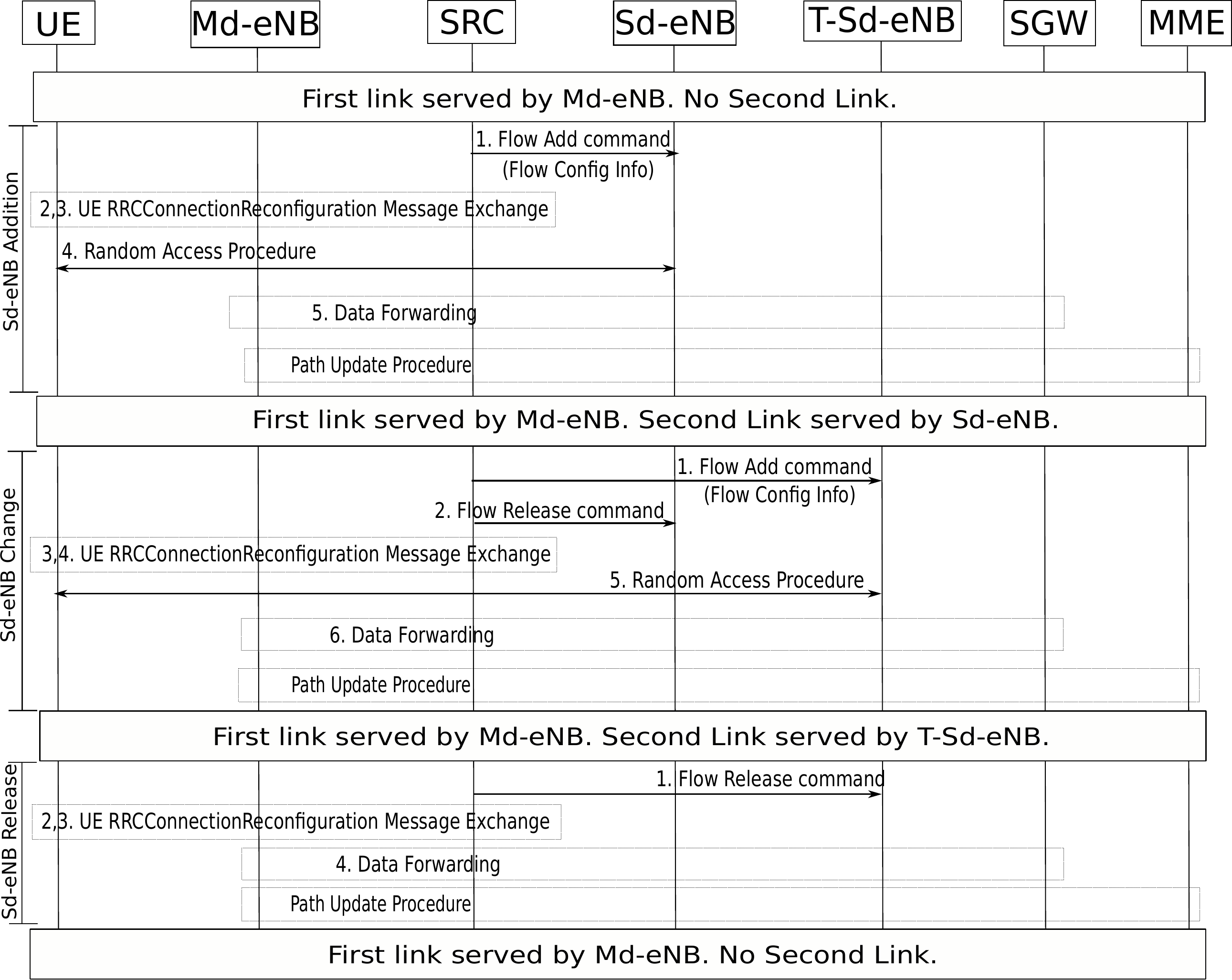}
	\caption{Signaling flow graph of Sd-eNB addition, Sd-eNB change and Sd-eNB release in proposed architecture.}
	\label{Flowgraph}
\end{figure}

The flow graphs for the various DC mobility events such as SeNB addition, SeNB change, and SeNB release in the legacy LTE architecture are presented in Section 10.1.2.8 of \cite{36300}. The procedures for these DC mobility events in the proposed architecture are displayed in Figure \ref{Flowgraph}. When a SC d-eNB (Sd-eNB) has to be added to a UE for DC, SRC sends an M-OF Flow Add command containing flow configuration information to the Sd-eNB. The RRCConnectionReconfiguration message exchanges then happen between the UE and SRC. UE then latches on to Sd-eNB using the random access procedure. The rest of the messages are the same as in the legacy SeNB addition procedure mentioned in \cite{36300}. Next, the figure shows the flow graph to change the second link from Sd-eNB to Target Sd-eNB (T-Sd-eNB). SRC sends an M-OF Flow Add command to T-Sd-eNB and an M-OF Flow Release command to Sd-eNB. Further procedures are the same as that explained for Sd-eNB addition. For the release of T-Sd-eNB, SRC sends an M-OF Flow Release command to Sd-eNB, and then a similar process is followed as that for Sd-eNB addition. \par
\begin{table}
	\caption{Number of signaling messages exchanged in the RAN during DC mobility events.}
	\begin{center}
		\begin{tabular}{|c|c|c|c|}
			\hline
			\textbf{Architecture } & \textbf{SeNB addition} & \textbf{SeNB change} & \textbf{SeNB release} \\
			\hline
		    Legacy & 3 & 5 & 2\\
			\hline
			Proposed & 1 & 2 & 1\\
			\hline
		\end{tabular}
	\end{center}
	\label{NumSignals}
\end{table}
The number of signaling exchanges in the DC mobility events of SeNB addition, SeNB change and SeNB release in legacy architecture and the proposed architecture are enumerated in Table \ref{NumSignals}. It presents the number of control signaling message exchanges between the network nodes in legacy LTE RAN and those between SRC and the data plane nodes in the proposed architecture. In the legacy LTE architecture, the number of control signaling exchanges for the addition of SeNB between MeNB and SeNB is 3 (SeNB addition request, SeNB addition request acknowledge and SeNB reconfiguration complete \cite{36300}). This has come down to 1 (Sd-eNB M-OF Flow add command) with the introduction of SRC in the proposed architecture (Figure \ref{Flowgraph}). A similar reduction in signaling messages can be shown in other DC operations as well. We demonstrate the effect of control signaling reduction in a heterogeneous network scenario.\par
We consider a 7-cell scenario with three hexagonal sectors in each cell and four pico cells per sector with overlapping coverage deployed in the central cell. Both macro and pico cells have a bandwidth of 10 MHz. The users are uniformly distributed in the center cell with 10 users per sector. We simulate different scenarios where the users move with speeds of 3kmph, 30kmph, and 60kmph, respectively. The users move within the sector with a fixed speed in a randomly chosen direction till a boundary is reached. Once they reach this boundary, they bounce back in a random direction such that they remain within a bounding box. The various characteristics of the network simulated are enumerated in Table \ref{Parameters}. We determine the number of DC mobility events, viz, secondary cell additions, secondary cell changes and secondary cell releases, experienced by a mobile user in a duration of 1000 secs. Events A2 and A4 \cite{36331} using RSRQ measurements are used as triggering events for these DC mobility events. The simulations are performed using NS-3 simulator \cite{NS3}.\par
\begin{table}
	\caption{Network Parameters.}
	\begin{center}
		\begin{tabular}{|l|l|}
			\hline
			\textbf{Parameter} &	\textbf{Value (Macro, Pico)} \\
			\hline
			Macro ISD, Pico radius & 500m, 50m\\
			\hline
			Transmit power & 46 dBm, 30 dBm \\
			\hline
			Antenna & Sectored, Omnidirectional \\
			\hline
			Antenna height & 32m, 10m \\
			\hline
			Path loss (d in km) & 128.1 + 37.6 log(d), 
			140.7 + 36.7 log(d) dB\\
			\hline
		\end{tabular}
	\end{center}
	\label{Parameters}
\end{table}
The simulation results obtained are illustrated in Table \ref{AvgMobilityEvents} for different speeds of users. These values, for instance, indicate that a UE traveling at a speed of 30kmph experiences an average of 0.02229 SeNB additions, 0.0289 SeNB changes and 0.0217 SeNB releases per second. The average number of message exchanges between eNBs in the legacy architecture and that between the SRC and d-eNBs in the proposed architecture are then calculated using Table \ref{NumSignals}.
\begin{table}
	\caption{Average number of DC mobility events per user per second.}
	\begin{center}
		\begin{tabular}{|c|c|c|c|}
			\hline
			\textbf{UE speed } & \textbf{3kmph} & \textbf{30kmph} & \textbf{60kmph} \\
			\hline
			\textbf{SeNB additions} & 0.00291 & 0.02229 & 0.04282 \\
			\hline
			\textbf{SeNB changes} & 0.00289 & 0.0289 & 0.05843 \\
			\hline
			\textbf{SeNB releases}  & 0.00233 & 0.0217 & 0.0423 \\
			\hline
		\end{tabular}
	\end{center}
	\label{AvgMobilityEvents}
\end{table}
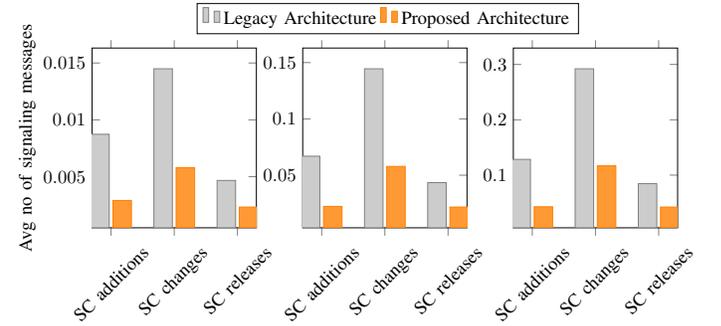
\begin{figure}[H]
	\centering
	\begin{tikzpicture}[scale=0.7]
	\pgfplotsset{
		group size=3,
		group gap=0.1cm,
		yticklabel style={/pgf/number format/fixed, /pgf/number format/precision=3},
		scaled y ticks=false,
	}
	
	\begin{axis}[
	scale=0.6,
	ybar,
	x=2cm,
	enlargelimits=0.15,
	legend style={at={(1.8,1.25)},
		anchor=north,legend columns=-1},
	ylabel={Avg no of signaling messages},
	symbolic x coords={SC additions, SC changes, SC releases},
	xtick=data,
	x tick label style={rotate=45}
	]
	\addplot [gray!60!gray,fill=gray!40!white] coordinates {(SC additions,0.00873) (SC changes,0.01449) (SC releases, 0.00467)};  
	\addplot [orange!20!orange,fill=orange!80!white] coordinates {(SC additions,0.002911) (SC changes,0.00579) (SC releases, 0.00233)}; 
	
	\legend{Legacy Architecture, Proposed Architecture}
	\end{axis}
	
	\begin{axis}[
	xshift=4cm,
	scale=0.6,
	ybar,
	x=2cm,
	enlargelimits=0.15,
	symbolic x coords={SC additions, SC changes, SC releases},
	xtick=data,
	x tick label style={rotate=45}
	]
	\addplot [gray!60!gray,fill=gray!40!white] coordinates {(SC additions,0.06687) (SC changes,0.1445) (SC releases, 0.0434)};   
	\addplot [orange!20!orange,fill=orange!80!white] coordinates {(SC additions,0.02229) (SC changes,0.0578) (SC releases, 0.0217)}; 
	
	\end{axis}
	
	\begin{axis}[
	xshift=8cm,
	scale=0.6,
	ybar,
	x=2cm,
	enlargelimits=0.15,
	symbolic x coords={SC additions, SC changes, SC releases},
	xtick=data,
	x tick label style={rotate=45}
	]
	\addplot [gray!60!gray,fill=gray!40!white] coordinates {(SC additions,0.12846) (SC changes,0.29216) (SC releases, 0.0846)};   
	\addplot [orange!20!orange,fill=orange!80!white] coordinates {(SC additions,0.04282) (SC changes,0.11686) (SC releases, 0.0423)}; 
	
	\end{axis}
	\end{tikzpicture}
	\caption{Amount of signaling messages exchanged in DC mobility events.}
	\label{Bargraph}
\end{figure}

Figure \ref{Bargraph} demonstrates the average number of signaling exchanges between the network nodes in the two architectures. We infer that there is a 66\% reduction in the signaling for SC addition in the proposed architecture. Similarly, there is 60\% and 50\% reduction in the signaling for SC change and SC release, respectively in the proposed architecture. Due to the reduction in control signaling messages between SRC and data plane nodes, the delay in initial connection setup for DC users is also reduced in the proposed architecture. 
\subsection{Improvement in System Performance}
\label{sec4.2}
In the proposed architecture, SRC contains the RRC functionality of all data plane entities. Therefore, all the signal strength measurements are received at the SRC, which then takes a centralized decision to select the DC users. We propose a centralized user association algorithm for DC in Algorithm \ref{alg1}. The key idea behind the algorithm is to dual connect those users whose secondary radio links are better than some of the existing primary radio links in a particular cell. The algorithm is executed periodically to take into account new user arrivals and departures in the system.\par
\begin{algorithm}
	\caption{Centralized DC algorithm.}\label{alg1}
	\begin{algorithmic}[1]
    	\State $\mathcal{B} \gets $ Set of all data plane nodes.
    	\State $\mathcal{U} \gets $ Set of UEs.
   		\State $count \gets 0$.
   		\State $A_1(\cdot) \gets 0$.
		\ForAll {$u \in \mathcal{U}$}
		\State $i^{\ast}\gets \arg\max\limits_{j \in \mathcal{B}} (RSRP(u,j))$ 
		\State $A_1(u)\gets i^{\ast}$
		\State Append $RSRP(u,i^{\ast})$ to $ RSRP_B(i^{\ast})$
		\EndFor
		\State $\mathcal{U} \gets $ UEs sorted according to their second best link
		\ForAll {$u \in \mathcal{U}$}	
		\While{$count < N$}
		\State $j^{\ast} \gets \arg\max\limits_{j \in \mathcal{B} \setminus {A_1(u)}} (RSRP(u,j))$ 
		\If	{$RSRP(u,j^{\ast})$ is better than at least $n$ existing links in $j^{\ast}$} 
		\State $A_2(u)\gets j^{\ast}$ 
		\State $count \gets count+1$
		\State Append $RSRP(u,j^{\ast})$ to $ RSRP_B(j^{\ast})$
		\EndIf
		\EndWhile		
		\EndFor
	\end{algorithmic}
\end{algorithm}
In Algorithm \ref{alg1}, $RSRP(a,b)$ denotes Reference Signal Received Power (RSRP) of user $a$ from a data plane network node $b$. For each user $u$, we determine the node $i^{\ast}$ which provides maximum RSRP (in step 6) and associate the user with it. We store this association in vector $A_1(\cdot)$. In step 8, we append this RSRP value to a vector corresponding to node $i^{\ast}$, denoted by $RSRP_B$. $RSRP_B(b)$ denotes a vector of RSRP values of users which are associated with node $b$. The users are now associated with the nodes which provide them with the best radio link. In Step 10, we sort all users in the order of the strength of their second best link. In the second iteration, we determine the node with the second best link for user $u$ (step 13). Only those users, whose second best link is better than at least $n$ existing links for that node, where $n$ is a configurable number, are selected for DC (step 14). The number of users that can be dual connected is limited to a fixed positive integer, $N$ (step 12). \par
We consider a 7-cell scenario with three hexagonal sectors in each cell and four pico cells per sector deployed in the central cell. The macro cell and pico cells operate at different carrier frequencies and have a bandwidth of 10 MHz each. The users are uniformly distributed in the middle cell and are stationary. We assume full buffer traffic for all users. The other parameters of the network simulated are listed in Table \ref{Parameters}. We compare the proposed algorithm with a distributed algorithm commonly used for DC, where a user with RSRP greater than a threshold ($RSRP_{th}$) for the secondary link is dual connected. We set the value of $RSRP_{th}$ to $-78$ dB. \par 
\definecolor{bblue}{HTML}{ACE2F6}
\definecolor{dblue}{HTML}{2C6EDA}
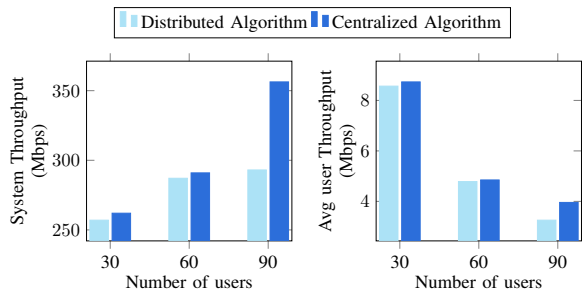
\begin{figure}[H]
	\centering
	\begin{tikzpicture}[scale=0.7]
	\pgfplotsset{
		group size=2,
		group gap=0.1cm,
	}
	\begin{axis}[
	scale=0.6,
	ybar,
	x=2.5cm,
	enlargelimits=0.15,
	legend style={at={(1.1,1.3)},
		anchor=north,legend columns=-1},
	y label style={at={(axis description cs:0.16,.5)},anchor=south,align=center},
	ylabel={System Throughput\\(Mbps)},
	symbolic x coords={30, 60, 90},
	xlabel={Number of users},
	xtick=data,
	]
	\addplot [bblue,fill=bblue] coordinates {(30,257.011125) (60,287.103) (90,293.0642)};  
	\addplot [dblue,fill=dblue] coordinates {(30,262.0188125) (60,290.9694) (90,356.305)}; 
	
	\legend{Distributed Algorithm, Centralized Algorithm}
	\end{axis}
	
	\begin{axis}[
	xshift=5.5cm,
	scale=0.6,
	ybar,
	x=2.5cm,
	enlargelimits=0.15,
	y label style={at={(axis description cs:0.25,.5)},anchor=south,align=center},
	ylabel={Avg user Throughput\\(Mbps)},
	symbolic x coords={30, 60, 90},
	xlabel={Number of users},
	xtick=data,
	]
	\addplot [bblue,fill=bblue] coordinates {(30,8.5670375) (60,4.78505) (90,3.2562688889) };  
	\addplot [dblue,fill=dblue] coordinates {(30,8.7339604167) (60,4.84949) (90,3.9589444) }; 
	
	\end{axis}
	\end{tikzpicture}
	\caption{Throughput performance of different algorithms.}
	\label{Algo_Thp}
\end{figure}
Figure \ref{Algo_Thp} illustrates the average system throughput and average UE throughput obtained by simulating the distributed algorithm and the proposed centralized algorithm for different user drops. In the proposed algorithm, SRC uses the knowledge of user measurements information gained through a global view of the network. The second link is added to the UE, only if its radio condition is better than at least one existing primary link in that cell. The simulation shows that the system, as well as average user performance, improves in the proposed architecture. Since the additional secondary links of DC users are added only when they are better than the existing primary links in that cell, the overall throughput of the system is improved.\par
\section{Conclusion}
\label{conc}
Due to diverse SRB establishment procedures across different DC modes, the control signaling in current DC architectures is inherently complex in nature. Also, the SN for a DC UE is selected based on a local view of the UE and the base station. To overcome these challenges in the design of the current architecture, we propose a multi-RAT RAN architecture based on SDN. Without any change in the CN, RAN is divided into two parts; a multi-RAT RAN controller with the associated protocol layers and the data plane entities with the user-plane protocol stack. This architecture provides unified control over the functionalities of RAN as well as a reduction in the control signaling in the network. A centralized algorithm for DC based on the proposed architecture is proposed. We show through simulations that it improves the system performance. 
\bibliographystyle{ieeetr}
\bibliography{IEEEabrv,SDN_paper5}
\end{document}